\documentclass[twocolumn,english,aps,prl,10pt,superscriptaddress]{revtex4}
\usepackage[T1]{fontenc}
\usepackage{amsmath,graphicx,amssymb,epsfig,babel,dsfont}

\renewcommand{\Im}{{\rm Im}}
\newcommand{\Tr}{{\rm Tr}}
\newcommand{\rd}{{\rm d}}

\begin{document}

\title{Giant thermal magnetoresistance in plasmonic structures}

\author{Ivan Latella}
\affiliation{Laboratoire Charles Fabry, UMR 8501, Institut d'Optique, CNRS, Universit\'{e} Paris-Saclay,
2 Avenue Augustin Fresnel, 91127 Palaiseau Cedex, France}

\author{Philippe Ben-Abdallah}
\email{pba@institutoptique.fr} 
\affiliation{Laboratoire Charles Fabry, UMR 8501, Institut d'Optique, CNRS, Universit\'{e} Paris-Saclay,
2 Avenue Augustin Fresnel, 91127 Palaiseau Cedex, France}
\affiliation{Universit\'{e} de Sherbrooke, Department of Mechanical Engineering, Sherbrooke, PQ J1K 2R1, Canada}

%\date{\today}

%\pacs{44.40.+a, 78.20.N-, 03.50.De, 66.70.-f}
\begin{abstract}
A giant thermal magnetoresistance is predicted for the electromagnetic transport of heat in magneto-optical plasmonic structures. In chains of InSb-Ag nanoparticles at room temperature, we found that the resistance can be increased by almost a factor 2 with magnetic fields of 2\,T. We show that this important change results from the strong spectral dependence of localized surface waves on the magnitude of the magnetic field. 
\end{abstract}

\maketitle

The ability to control electromagnetic energy in deep subwavelength volumes with surface plasmons or phonon polaritons has attracted much attention in nanoscale science during the last decade. In the region where localized surface waves exist, the electromagnetic field is coupled to atoms and charges oscillations. Such vibrations are responsible for a strong dissipation by Joule effect and the dissipated heat spread out through the structure both by radiation and conduction. This emerging branch of plasmonics is called thermoplasmonics~\cite{Govorov,Baffou1,Baffou2,Baffou3}. Original mechanisms for the heat transport  have been described in these media such as anomalous regimes~\cite{pba1} in plasmonic networks due to long range interactions as well as amplification phenomena for photon heat tunneling~\cite{pba2} due to cooperative effects. 
Besides these fundamental results, the control of temperature fields at subwavelength scales has found several important applications in medical therapy~\cite{Han,Loo,Jain}, heat-assisted magnetic recording~\cite{Challener,Stipe}, chemical catalysis~\cite{Christopher}, thermotronics~\cite{Baowen,pba3,pba4}, thermal lithography~\cite{Kim}, and thermophotovoltaics energy conversion~\cite{DiMatteo,Arvin}. 
Today, the dynamic control of heat transport is undoubtedly a new frontier in thermoplasmonics.  
A step forward in this direction has been recently taken with the discovery of a photon thermal Hall effect~\cite{pba_Hall} in magneto-optical nanoparticle lattices in presence of an external magnetic field. This thermomagnetic effect consists in the appearance of a radiative heat flux transversally to the direction of a primary temperature gradient because of a local symmetry breaking induced by the magnetic field.

In the present Letter we predict another thermomagnetic effect: a magnetoresistance for the  heat flux carried by thermal photons. In the ordinary magnetoresistance~\cite{Thomson}, the change in material's resistivity with the application of a magnetic field is relatively small, and only differences of a few percent have been observed at room temperature. On the other hand, giant magnetoresistance (GMR) effects of up to 50\% have been reported~\cite{Baibich} at low temperature  in magnetic-nonmagnetic multilayer structures. Here we anticipate a thermal analog of this effect at room temperature in magneto-optical lattices with reasonably small magnetic fields.  

In the matter at hand we need to characterize the heat flux exchanged by thermal radiation in magneto-optical many-body systems out of thermal equilibrium. Lorentz reciprocity is broken in these systems, so that permanent heat currents can exist even at thermal equilibrium~\cite{Fan}. If we consider a set of $N$ subwavelength media at arbitrary temperatures $T_i$ immersed in a thermal bath at a different temperature, such media can exchange electromagnetic energy between them as well as with the bath. The associated heat transfer can be described using a Landauer-like formalism~\cite{pba5,Riccardo,Nikbakht,Incardone}, from which the net heat flux exchanged between two nonreciprocal media $i$ and $j$ can be written under the general form
\begin{eqnarray}
&&\varphi_{ij}(\mathbf{H})= \label{Eq:InterpartHeatFlux}\\
&&\qquad\int_{0}^{\infty}\frac{\rd\omega}{2\pi}\,[\Theta(\omega,T_{i})\mathcal{T}_{i,j}(\omega,\mathbf{H})-\Theta(\omega,T_{j})\mathcal{T}_{j,i}(\omega,\mathbf{H})],\nonumber
\end{eqnarray}
where $\Theta(\omega,T)={\hbar\omega}/[{e^{\frac{\hbar\omega}{k_B T}}-1}]$ is the mean energy of a harmonic oscillator in
thermal equilibrium at temperature $T$ and $\mathcal{T}_{i,j}(\omega,\mathbf{H})$ denotes the transmission coefficients at the frequency $\omega$ under the action of a magnetic field $\mathbf{H}$.
These transmission coefficients are defined as~\cite{Nikbakht}
\begin{equation}
  \mathcal{T}_{i,j}(\omega,\mathbf{H})=2\Im\Tr\bigl[\mathds{A}_{ij}\Im\bar{\bar{\boldsymbol{\chi}}}_j\mathds{C}_{ij}^{\dagger}\bigr],
\end{equation}
where the susceptibility tensor $\bar{\bar{\boldsymbol{\chi}}}_j$ and the matrices $\mathds{A}_{ij}$ and $\mathds{C}_{ij}$ are given by~\cite{Nikbakht} 
\begin{eqnarray}
&&\bar{\bar{\boldsymbol{\chi}}}_j(\mathbf{H})=\bar{\bar{\boldsymbol{\alpha}}}_{j}(\mathbf{H})-i\frac{k^3}{6\pi} \bar{\bar{\boldsymbol{\alpha}}}_{j}(\mathbf{H})\bar{\bar{\boldsymbol{\alpha}}}_{j}^{\dagger}(\mathbf{H}),\\
&&\mathds{A}_{ij}=\left[\mathds{1}-k^2\hat{\mathds{\alpha}}\mathds{B}\right]_{ij}^{-1},\\
&&\mathds{C}_{ij}=k^2\mathds{H}_{ik}\mathds{A}_{kj}.
\end{eqnarray}
Here $\mathds{B}_{ij}=(1-\delta_{ij})\mathds{G}_{ij}^0$ and $\mathds{H}_{lm}=i\frac{k}{6\pi}\delta_{lm}\mathds{1}+\mathds{B}_{lm}$, while $\bar{\bar{\boldsymbol{\alpha}}}_{j}$ is the polarizability tensor of the $j^{th}$ object, $\hat{\mathds{\alpha}}=\text{diag}(\bar{\bar{\boldsymbol{\alpha}}}_{1},\dots,\bar{\bar{\boldsymbol{\alpha}}}_{N})$ is the polarizability matrix, and $\mathds{G}_{ij}^0=\frac{\exp({\rm i}kr_{ij})}{4\pi r_{ij}}\left[\left(1+\frac{{\rm i}kr_{ij}-1}{k^{2}r_{ij}^{2}}\right)\mathds{1}+\frac{3-3{\rm i}kr_{ij}-k^{2}r_{ij}^{2}}{k^{2}r_{ij}^{2}}\widehat{\mathbf{r}}_{ij}\otimes\widehat{\mathbf{r}}_{ij}\right]$
is the free space Green tensor
($\widehat{\mathbf{r}}_{ij}\equiv\mathbf{r}_{ij}/r_{ij}$, $\mathbf{r}_{ij}$ being the vector linking the center of
dipoles $i$ and $j$, $r_{ij}=|\mathbf{r}_{ij}|$, and $\mathds{1}$ stands for the unit dyadic tensor).

When the system is at equilibrium at temperature, let say, $T_j$, it follows from the general expression (\ref{Eq:InterpartHeatFlux}) that the media $i$ and $j$ still exchange an energy flux
\begin{equation}
\varphi^{eq}_{ij}(\mathbf{H})=\int_{0}^{\infty}\frac{\rd\omega}{2\pi}\,\Theta(\omega,T_j)[\mathcal{T}_{i,j}(\omega,\mathbf{H})-\mathcal{T}_{j,i}(\omega,\mathbf{H})]\label{Eq:equilibriumHeatFlux}.
\end{equation}
This flux corresponds to the permanent current recently highlighted by Zhu and Fan~\cite{Fan} in a system of three magneto-optical particles, which is a function of the asymmetry degree in the system.  
If the transmission coefficients $\mathcal{T}_{i,j}$ and $\mathcal{T}_{j,i}$ are not equal  in nonreciprocal systems, they are nonetheless not independent ones from the other. Indeed, when we consider a system at equilibrium at temperature $T$, the net flux received by each medium must vanish and therefore, the relation
\begin{eqnarray}
&&\sum_j \varphi_{ij}(\mathbf{H})= \label{Eq:netHeatFlux}\\
&&\qquad\int_{0}^{\infty}\frac{\rd\omega}{2\pi}\,\Theta(\omega,T) \sum_j[\mathcal{T}_{i,j}(\omega,\mathbf{H})-\mathcal{T}_{j,i}(\omega,\mathbf{H})]=0\nonumber
\end{eqnarray}
holds for any $i=\overline{1,N}$. Since this flux vanishes whatever is the equilibrium temperature, the following general relation
\begin{equation}
 \underset{j}{\sum}[\mathcal{T}_{i,j}(\omega,\mathbf{H})-\mathcal{T}_{j,i}(\omega,\mathbf{H})]=0\label{Eq:relation_coeffs}
\end{equation}
 must be satisfied between the transmission coefficients. 
This condition is remarkable because it assures that the total flux received by each medium in the many-body system vanishes at equilibrium despite of the presence of an equilibrium flux. As a consequence, a body in the system cannot cool down or heat up under these conditions, and the second law of thermodynamics is not violated by the occurrence of the permanent flux.
In the particular case of two-body systems, this relation immediately leads to
\begin{equation}
\mathcal{T}_{1,2}(\omega,\mathbf{H})=\mathcal{T}_{2,1}(\omega,\mathbf{H}),\label{Eq:coeffs-2body}
\end{equation}
so that the energy transmission coefficients become symmetric regardless of the nonreciprocity.

\begin{figure}%[!hbt]
%\centering
\includegraphics[angle=0,scale=0.3]{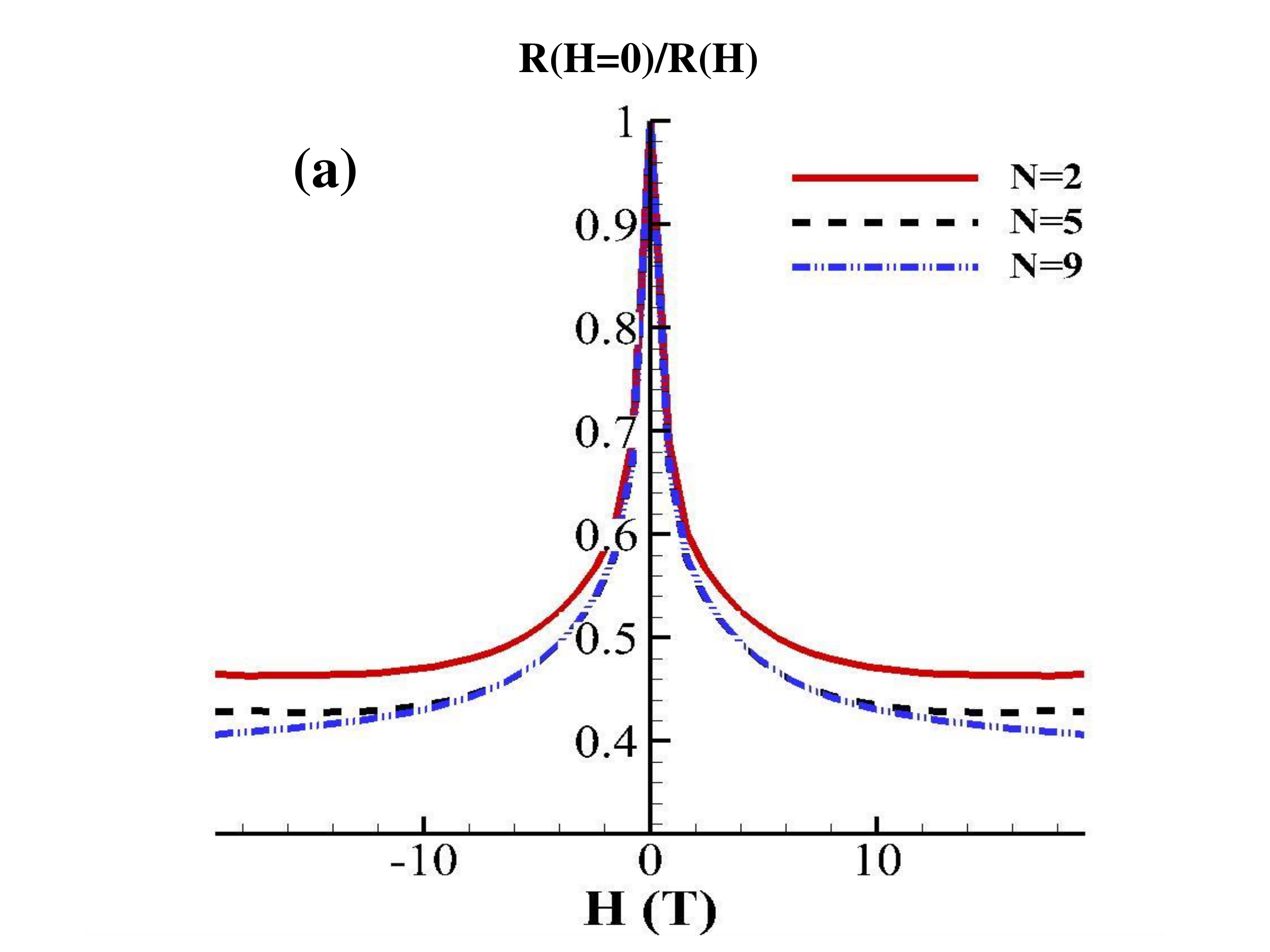}
\includegraphics[angle=0,scale=0.3]{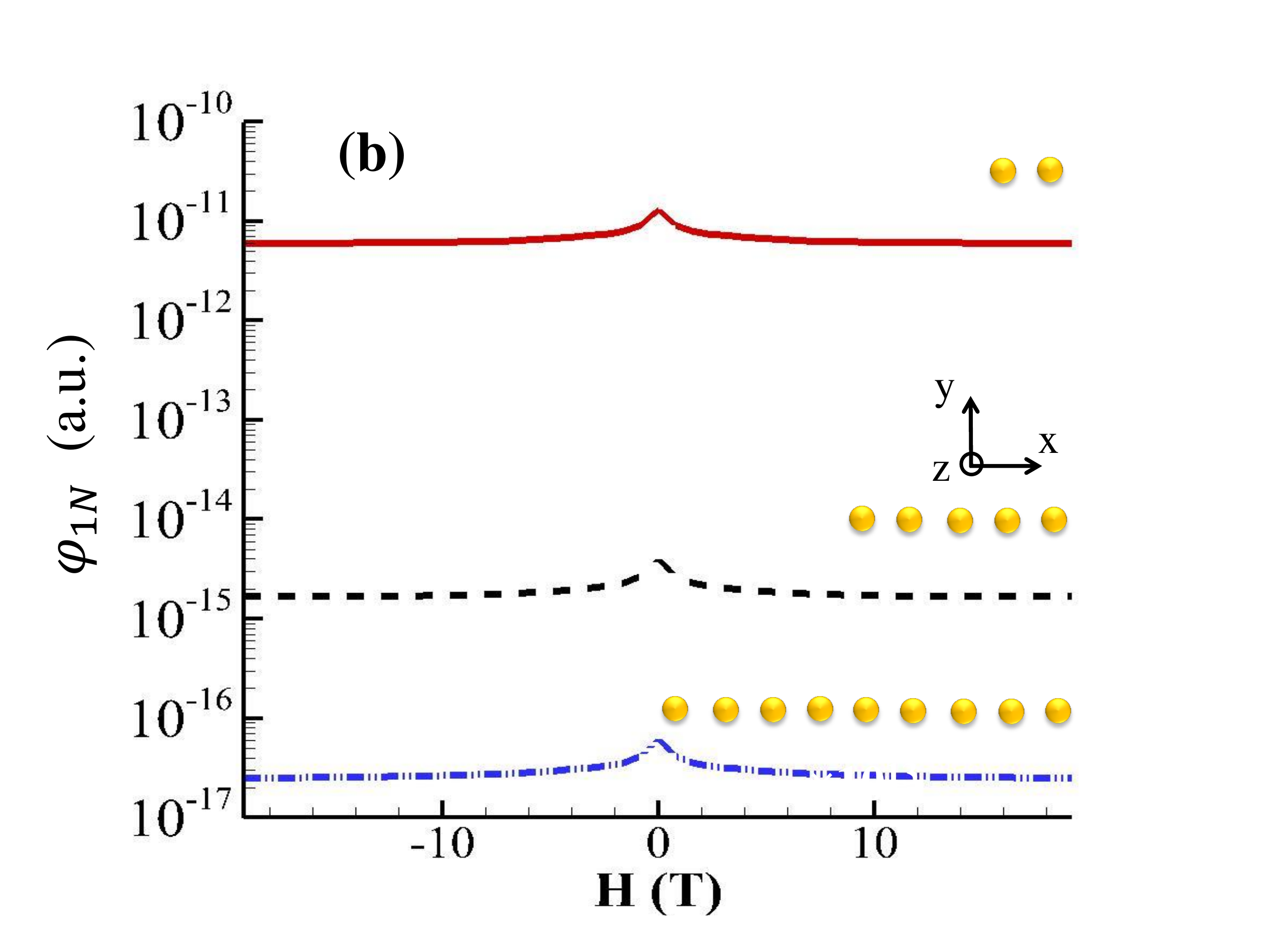}
\caption{(a) Thermal magnetoresistance of InSb linear nanoparticle chains of different length at $T=300\,$K as a function of the strength of an external magnetic field $H$ orthogonal to the chain. 
The positive sign for the strength of the field corresponds to a field pointing outwards and the negative sign to a field in the opposite direction. The radius of the nanoparticles is $r=100\,$nm and the separation distance between two adjacent particles is $d=2r$ (edge to edge). 
(b) Net energy flux exchanged along the chain between the first and the last particle as a function of the external magnetic field (orthogonal to the chain). A temperature difference $\Delta T=50\,$K is applied at $T=300\,$K.} 
\label{Fig_1}
\end{figure} 

Furthermore, from expressions (\ref{Eq:InterpartHeatFlux}) and (\ref{Eq:equilibriumHeatFlux}) one can deduce the nonequilibrium contribution to the total heat flux 
\begin{equation}
\varphi^{neq}_{ij}(\mathbf{H})=\int_{0}^{\infty}\frac{\rd\omega}{2\pi}\,[\Theta(\omega,T_i)-\Theta(\omega,T_j)]\mathcal{T}_{i,j}(\omega,\mathbf{H}),\label{Eq:nonequilibriumHeatFlux}
\end{equation}
which depends on the difference of distribution functions of the media. In the configurations analyzed below, this contribution strongly dominates over the equilibrium one~\cite{SupplMat}. Moreover, this expression of $\varphi^{neq}_{ij}$ allows us to define a thermal conductance $G_{ij}(\mathbf{H})=\lim_{|\Delta T|\to0} \varphi^{neq}_{ij}(\mathbf{H})/|\Delta T|$ in non-reciprocal systems which is consistent with the classical definition in reciprocal systems, where $\Delta T=T_i-T_j$. The inverse of this conductance is the thermal magnetoresistance
\begin{equation}
R_{ij}(\mathbf{H})=\left(\int_{0}^{\infty}\frac{\rd\omega}{2\pi}\frac{\partial\Theta}{\partial T}\mathcal{T}_{i,j}(\omega,\mathbf{H})\right)^{-1}.\label{Eq:neq-resistance}
\end{equation}

\begin{figure}%[!hbt]
%\centering
\includegraphics[angle=0,scale=0.3,angle=0]{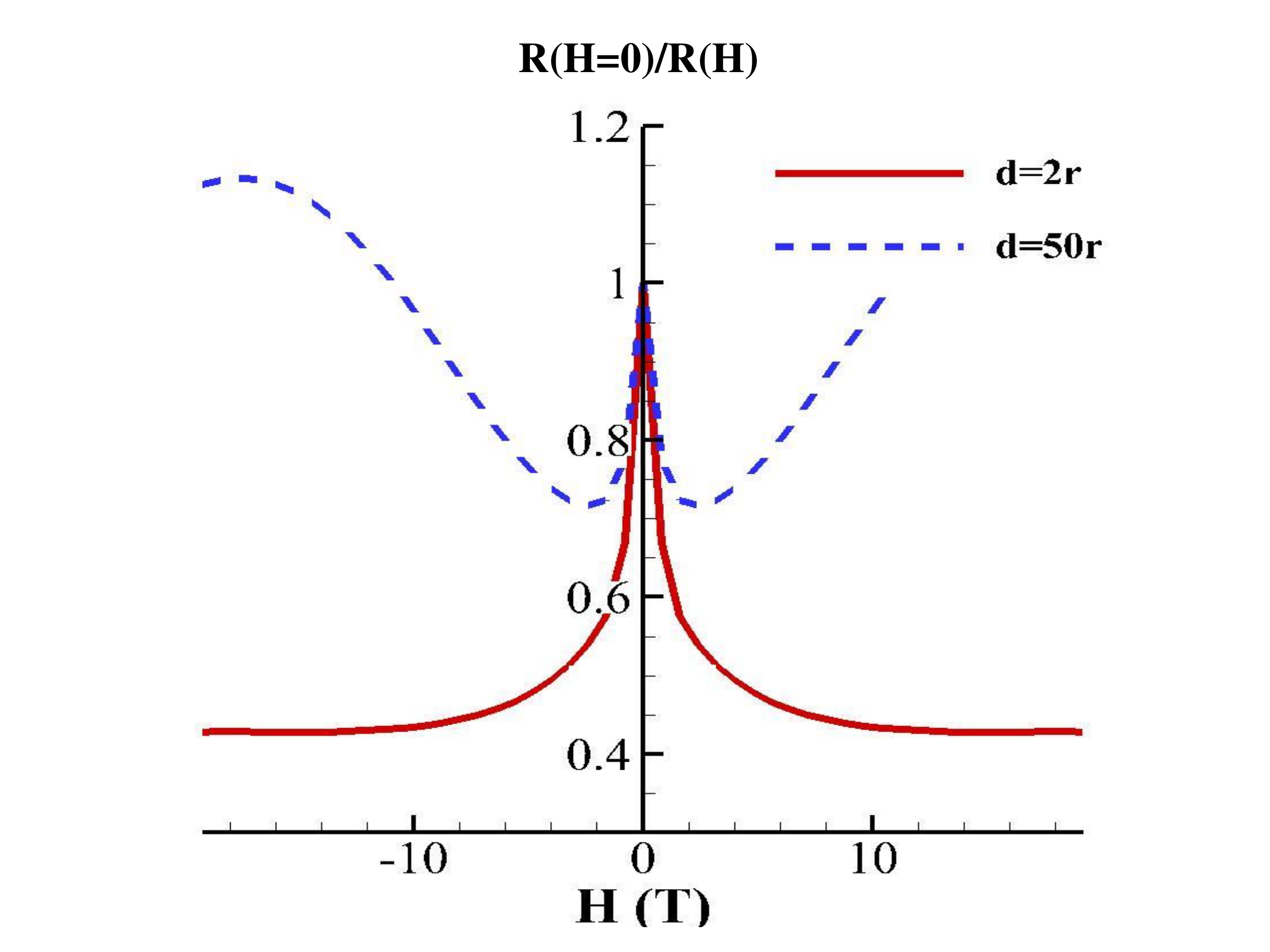}
\caption{Thermal magnetoresistance of InSb linear nanoparticle chains with respect to the strength of an external magnetic field $H$ at $T=300\,$K. The radius and number of nanoparticles are $r=100\,$nm and $N=5$, respectively, for  separation distances $d=2r$ (in the near field) and $d=50r$ (corresponding to the far field).}
\label{Fig_2}
\end{figure}

\begin{figure}%[!hbt]
%\centering
\includegraphics[angle=0,scale=0.3]{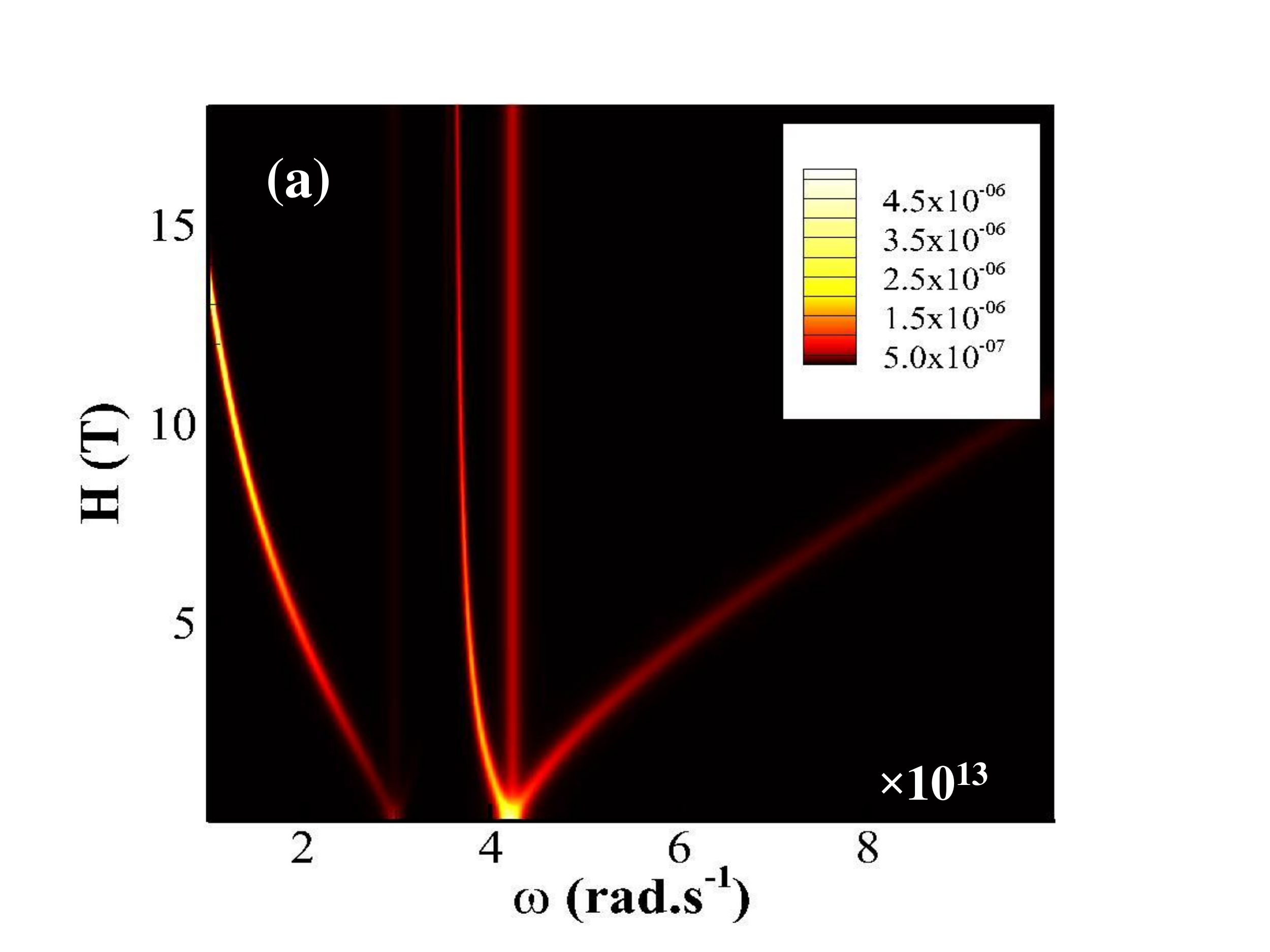}
\includegraphics[angle=0,scale=0.3]{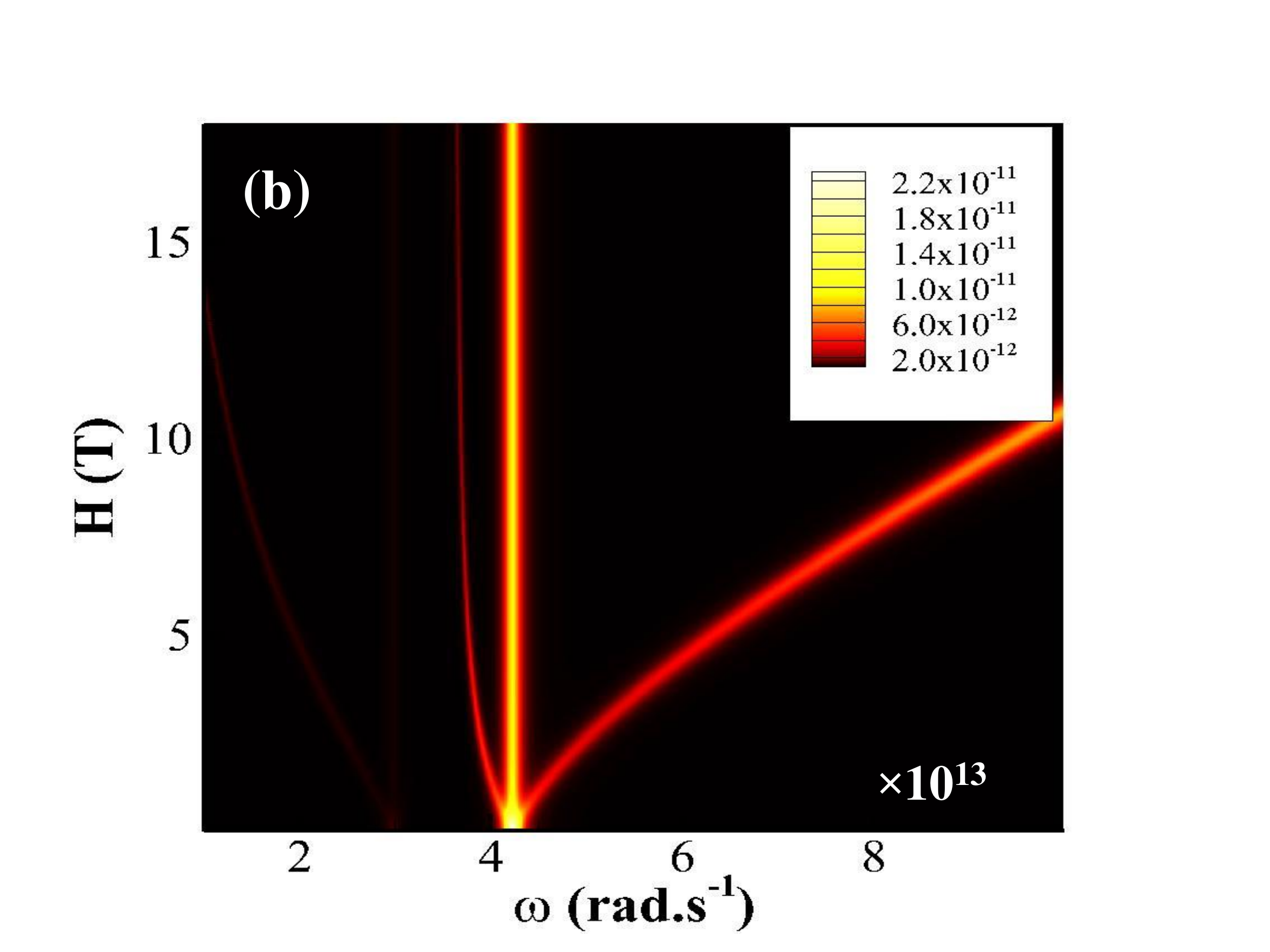}
\includegraphics[angle=0,scale=0.3]{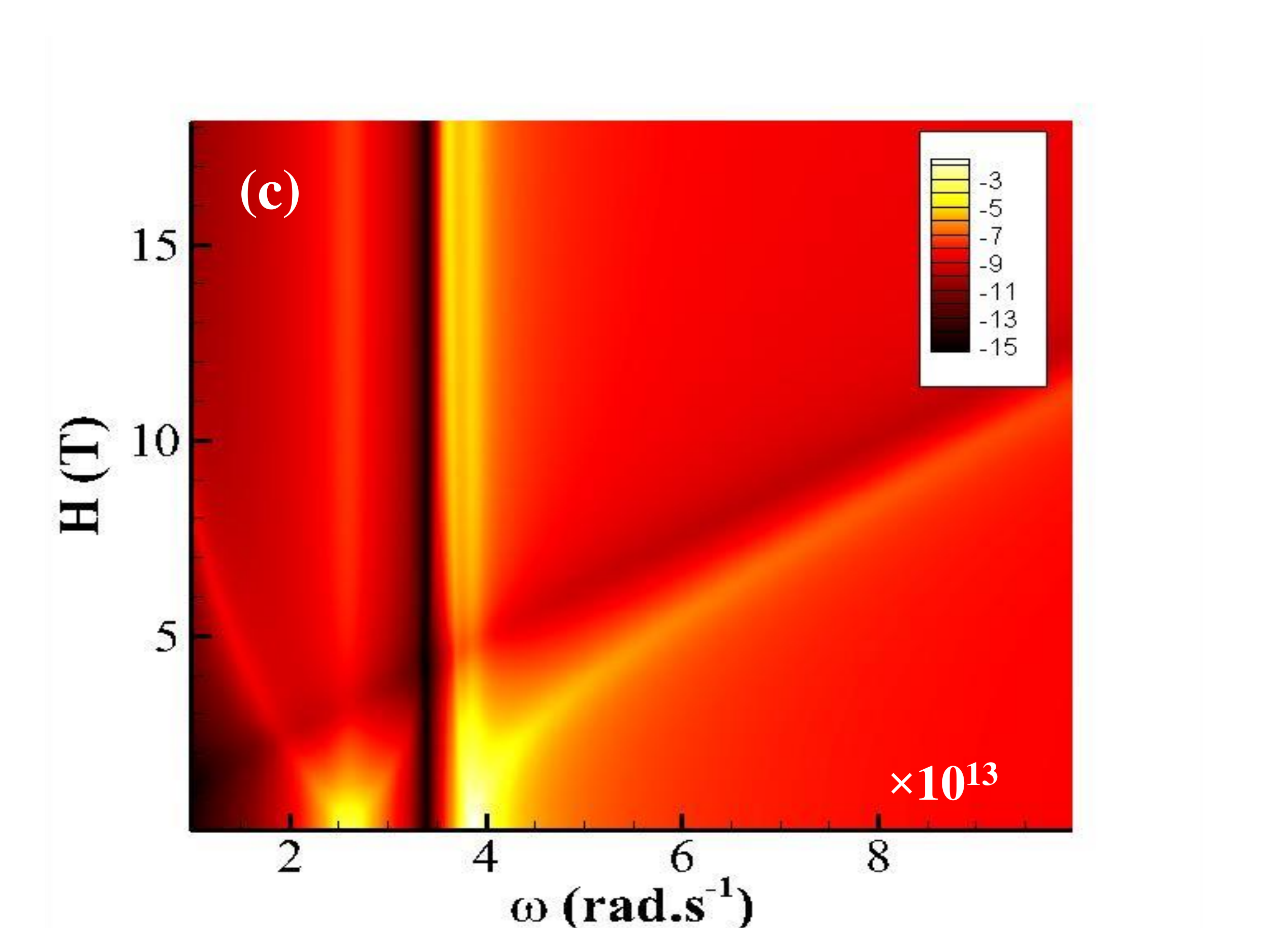}
\caption{(a)-(b) Energy transmission coefficients $\mathcal{T}_{1,N}$ in the $(\omega,H)$ plane for a linear chain of $N=5$ InSb-particles of radius $r=100\,$nm. The separation distance (edge to edge) between the particles is $d=2r$ in (a) and $d=50r$ in (b). (c) Location of resonances of the InSb particles shown in the $(\omega,H)$ plane. The plot shows the function $f(\omega,H)=\ln\left(|(\varepsilon_3+2\varepsilon_h)[(\varepsilon_1+2\varepsilon_h)^2-\varepsilon^2_2]|^{-1}\right)$.}
\label{Fig_3}
\end{figure}

To investigate the behavior of the thermal magnetoresistance in a concrete situation, we consider a linear chain of $N$ magneto-optical nanoparticles immersed in an external magnetic field of strength $H$ orthogonal to the axis of the chain (the anisotropic magnetoresistance is not considered here). The resistance of the chain is then given by $R(H)\equiv R_{1N}(\mathbf{H})$. Moreover, to quantify the influence of the magnetic field we compute the ratio $R(0)/R(H)$ of the resistance  without field to its value at the given field.
In particular, we first consider linear chains of $n$-doped InSb nanoparticles with different lengths.
Under these conditions, the permittivity tensor of the particles takes the form~\cite{Moncada}
\begin{equation}
\bar{\bar{\boldsymbol{\varepsilon}}}=\left(\begin{array}{ccc}
\varepsilon_{1} & -i\varepsilon_{2} & 0\\
i\varepsilon_{2} & \varepsilon_{1} & 0\\
0 & 0 & \varepsilon_{3}
\end{array}\right),\label{Eq:permittivity}
\end{equation}
with
\begin{eqnarray}
\varepsilon_{1}(H)\!\!&=&\!\!\varepsilon_\infty\bigg(1+\frac{\omega_L^2-\omega_T^2}{\omega_T^2-\omega^2-i\Gamma\omega}+\frac{\omega_p^2(\omega+i\gamma)}{\omega[\omega_c^2-(\omega+i\gamma)^2]}\bigg) \label{Eq:permittivity1},\nonumber\\
\varepsilon_{2}(H)\!\! &=& \!\!\frac{\varepsilon_\infty\omega_p^2\omega_c}{\omega[(\omega+i\gamma)^2-\omega_c^2]}\label{Eq:permittivity2},\nonumber\\
\varepsilon_{3} \!\!&=&\!\! \varepsilon_\infty\bigg(1+\frac{\omega_L^2-\omega_T^2}{\omega_T^2-\omega^2-i\Gamma\omega}-\frac{\omega_p^2}{\omega(\omega+i\gamma)}\bigg)\label{Eq:permittivity3}.
\end{eqnarray}
Here $\varepsilon_\infty=15.7$ is the infinite-frequency dielectric constant, $\omega_L=3.62\times10^{13}\,\mathrm{rad\,s}^{-1}$ is the longitudinal optical phonon frequency, $\omega_T=3.39\times10^{13}\,\mathrm{rad\,s}^{-1}$ is the transverse optical phonon frequency, $\omega_p=(\frac{ne^2}{m^*\varepsilon_0\varepsilon_\infty})^{1/2}$ is the plasma frequency of free carriers of density $n=1.68\times10^{17}\,\mathrm{cm}^{-3}$, charge $e$, and effective mass $m^*=1.99\times 10^{-32}\,\mathrm{kg}$, $\varepsilon_0$ is the vacuum permittivity, $\Gamma=5.65\times10^{11}\,\mathrm{rad\,s}^{-1}$ is the phonon damping constant,$\gamma=3.39\times10^{12}\,\mathrm{rad\,s}^{-1}$ is the free carrier damping constant, and $\omega_c=eH/m^*$ is the cyclotron frequency. Thus, the polarizability tensor for a spherical particles can be described, including the radiative corrections, by the following anisotropic polarizability~\cite{Albaladejo}
\begin{equation}
\bar{\bar{\boldsymbol{\alpha}}}_{i}(\omega)=\left( \bar{\bar{\boldsymbol{1}}}-i\frac{k^3}{6\pi} \bar{\bar{\boldsymbol{\alpha_0}}}_{i}\right)^{-1} \bar{\bar{\boldsymbol{\alpha_0}}}_{i}\label{Eq:Polarizability},
\end{equation}
where $ \bar{\bar{\boldsymbol{\alpha_0}}}_{i}$ denotes the quasistatic polarizability of the $i^{th}$ particle and $k=\omega/c$, $c$ being the speed of light in vacuum. For spheres made with magneto-optical materials and which are embedded inside an isotropic host of permittivity $\varepsilon_h$, this quasistatic polarizability reads
\begin{equation}
 \bar{\bar{\boldsymbol{\alpha_0}}}_{i}(\omega)=4\pi r^3\big(\bar{\bar{\boldsymbol{\varepsilon}}}-\varepsilon_h\bar{\bar{\boldsymbol{1}}}\big)\big(\bar{\bar{\boldsymbol{\varepsilon}}}+2\varepsilon_h\bar{\bar{\boldsymbol{1}}}\big)^{-1}\label{Eq:Polarizability2},
\end{equation}
where $r$ is the radius of the particles.
Examples of thermal magnetoresistance ratio $R(0)/R(H)$ at $T=300\,$K for these particles are shown in Fig.~\ref{Fig_1}(a) for different chain lengths. In Fig.~\ref{Fig_1}(b) we also show for this case the net flux $\varphi_{1N}(H)$ exchanged between the first and the $N^{th}$ particle when a temperature difference $\Delta T=50\,$K is applied around an equilibrium temperature $T=300\,$K. We observe that the resistance of the chain significantly increases with a magnetic field applied perpendicularly to the chain axis. After an abrupt change, the resistance reaches approximately a constant value. For very large field strengths (not shown) the resistance can again undergo considerable changes. We emphasize that the result presented in Fig.~\ref{Fig_1}(a) indicates a drastic variation of thermal magnetoresistance over a relatively small range of field strength. Whatever is the length of the chain, we observe that the resistance increment is about 70\% with a magnetic field of approximately 2\,T and is larger for more intense fields. Furthermore, Fig.~\ref{Fig_2} describes how the thermal magnetoristance evolves when the separation distance $d$ between the particles is modified from the regime of near-field interaction ($d=2r$) to the regime of far-field interaction ($d=50r$). In both cases, the resistance increases for weak strengths, somewhat larger than $H=0$. On the other hand, as the magnetic field strength increases the resistance decreases in dense chains while it increases in dilute ones before reaching a plateau. 

\begin{figure}%[!hbt]
%\centering
\includegraphics[angle=0,scale=0.3]{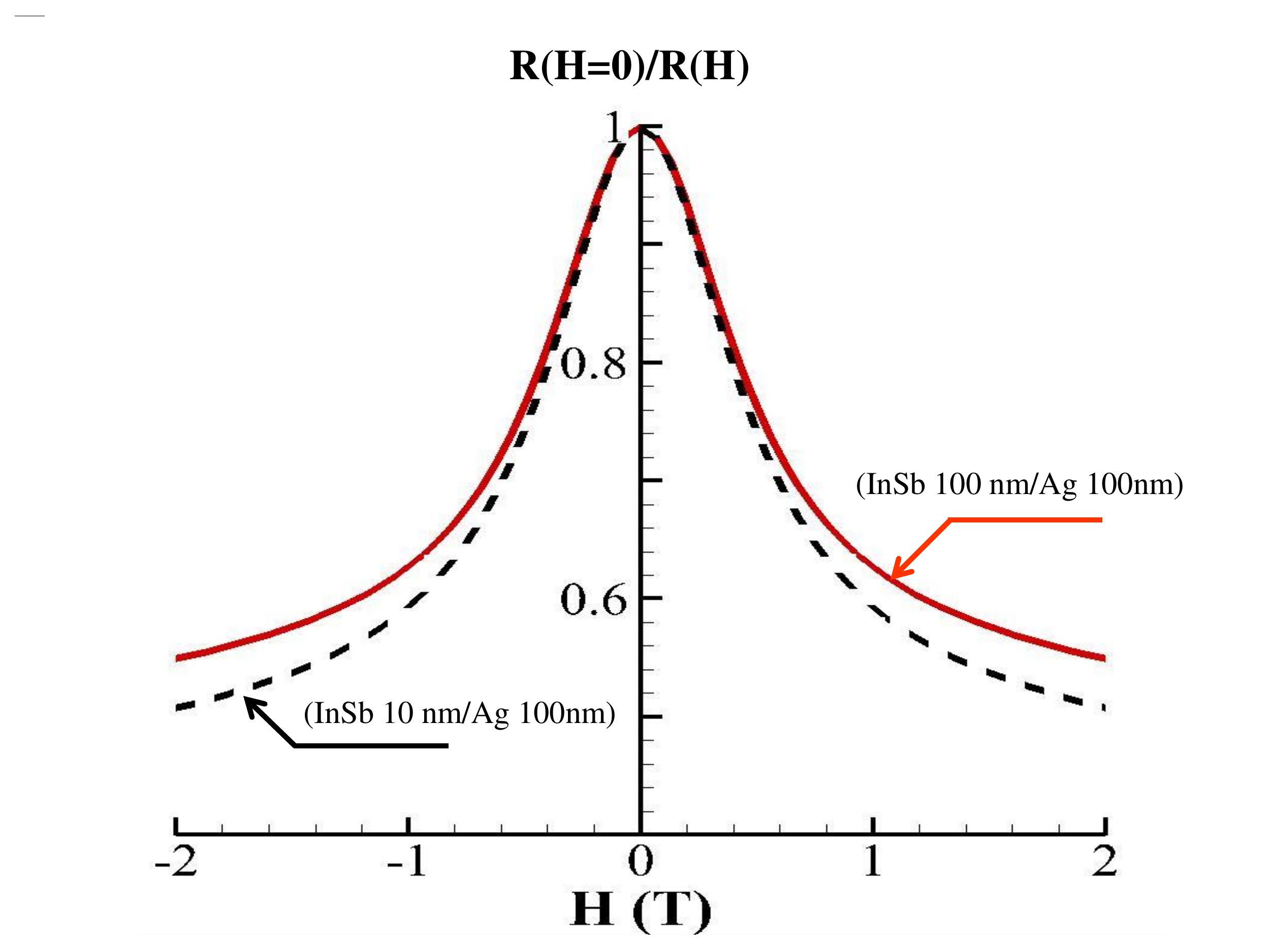}
\caption{Thermal magnetoresistance of InSb-Ag binary chains of $N=5$ particles at $T=300\,$K with different radii. The separation distance (edge to edge) is the sum of the two radii.}
\label{Fig_4}
\end{figure}

To give insight into the physical mechanism behind this behavior, we show in Fig.~\ref{Fig_3} the energy transmission coefficients along InSb nanoparticle chains in the frequency-magnetic field domain $(\omega,H)$. It can be seen that mainly three branches play a relevant role in the heat transfer along the chain. One of these branches is independent of the magnetic field, whereas the other two exhibit a marked dependence on the strength of the field. A comparison of these plots with a mapping of the particle resonances in the same domain demonstrates that the channels for heat exchanges along the chain are the associated localized resonances~\cite{SupplMat}. As shown in Fig.~\ref{Fig_3}(c), these channels strongly depend on the strength of the magnetic field. In addition, since Wien's frequency is shifted toward higher or lower frequency when the temperature is changed, these magnetic-dependent channels participate more or less to the heat transfer. The comparison of Fig.~\ref{Fig_3}(a) and Fig.~\ref{Fig_3}(b) with Fig.~\ref{Fig_3}(c) reveals that the large variations of thermal magnetoresistance originate from the spectral shift of resonant modes with respect to the field strength. Finally, we point out that the magnetoresistive effect can be larger in more complex structures. As can be seen in Fig.~\ref{Fig_4}, in binary InSb-Ag nanopaticle chains the increment in the resistance is even more noticeable.

We now briefly discuss the potential of this giant thermal magnetoresistance (GTMR) in some practical applications. In the linear regime, the heat flux between two elements inside a nanoparticle network takes the general form
\begin{equation}
\varphi\approx(\alpha H+\beta)\Delta T\label{Eq:HeatFlux_sensor}.
\end{equation}
Therefore, by locally measuring the temperature difference $\Delta T$ at the given field $H$, the heat flux propapgating in the network can be locally evaluated. Reversely, the GTMR can be exploited to make a purely thermal measurement of the magnetic field intensity. In addition, the GTMR could be implemented in a strain gauge to make local measurements of the thermal expansion by identifying the affine parameters in expression (\ref{Eq:HeatFlux_sensor}) for a given magnetic field.

In summary, we have shown the existence of a GTMR in simple InSb and binary InSb-Ag nanopaticle chains. The resistance is increased by almost a factor of 2 when a magnetic field of few teslas is applied orthogonally to the chains. This GTMR results from a strong spectral shift of localized surface waves supported by the particles under the action of a magnetic field. This effect is promising for practical applications, especially in the field of thermal management at nanoscale as well as for magnetic sensing with temperature or heat flux measurements.
We have limited this study to simple InSb or InSb-Ag chains where the GTMR requires the application of relatively strong magnetic field. But a giant thermal magnetoresistance could be observable with weaker fields using different materials or combining different magneto-optical and nonmagnetic materials.

\begin{acknowledgments}
P.B.-A. acknowledges discussions with Profs. Qi Hong, Zhao Junming and Jia Zi-Xun. 
\end{acknowledgments}

\end{document}